\documentclass[final]{svjour3}
\usepackage[dvips]{graphicx}
\usepackage{rotating}
\usepackage{amssymb}
\usepackage{mathptmx}
\usepackage[numbers]{natbib}
\makeatletter
\journalname{Journal of Low Temperature Physics}

\bibpunct{}{}{,}{s}{}{,}

\def\up{\uparrow}
\def\down{\downarrow}
\def\mbf#1{\mathbf#1}
\def\lesssim{\ \raise.3ex\hbox{$<$}\kern-0.8em\lower.7ex\hbox{$\sim$}\ }
\def\gesim{\ \raise.3ex\hbox{$>$}\kern-0.8em\lower.7ex\hbox{$\sim$}\ }
\def\s{\sigma}
\begin{document}
\newcommand{\hdblarrow}{H\makebox[0.9ex][l]{$\downdownarrows$}-}
\title{Spin susceptibility and effects of inhomogeneous strong pairing fluctuations in a trapped ultracold Fermi gas}
\author{H. Tajima \and R. Hanai \and Y. Ohashi}
\institute{Department of Physics, Faculty of Science and Technology, Keio University,\\ 3-14-1, Hiyoshi, Kohoku-ku, Yokohama 223-8522, Japan\\
Tel.:+81-45-566-1454 Fax:+81-45-566-1672
\email{htajima@rk.phys.keio.ac.jp}
}
\date{\today}
\maketitle
\begin{abstract}
We theoretically investigate magnetic properties of a unitary Fermi gas in a harmonic trap. Including strong pairing fluctuations within the framework of an extended $T$-matrix approximation (ETMA), as well as effects of a trap potential within the local density approximation (LDA), we calculate the local spin susceptibility $\chi(T,r)$ above the superfluid phase transition temperature $T_{\rm c}$. We show that the formation of preformed singlet Cooper pairs anomalously suppresses $\chi(T,r)$ in the trap center near $T_{\rm c}$. We also point out that, in the unitarity limit, the spin-gap temperature in a uniform Fermi gas can be evaluated from the observation of the spatial variation of $\chi(T,r)$. Since a real ultracold Fermi gas is always in a trap potential, our results would be useful for the study of how this spatial inhomogeneity affects thermodynamic properties of an ultracold Fermi gas in the BCS-BEC crossover region.

\end{abstract}
\section{Introduction}
An ultracold Fermi gas provides us the unique opportunity that we can systematically study physical properties of a many fermion system at various interaction strengths, by adjusting the threshold energy of a Feshbach resonance\cite{vgurarie07,sgiorgini08,ibloch08,cchin10}. Indeed, by using this advantage, the so-called BCS (Bardeen-Cooper-Shrieffer)-BEC (Bose-Einstein condensation) crossover\cite{dmeagles69,NSR,SadeMelo,rhaussmann94,yohashi02} has experimentally been realized in $^{40}$K\cite{caregal04} and $^6$Li\cite{mwzwierlein04,Kinast,Bartenstein} Fermi gases, where a BCS-type Fermi superfluid continuously changes into the BEC of tightly bound molecules, with increasing the strength of a pairing interaction. In this sense, we can now deal with a Fermi superfluid and a Bose superfluid in a unified manner.
\par
Recently, the spin susceptibility has become accessible in the BCS-BEC crossover regime of an ultracold Fermi gas\cite{Sanner,Sommer,yrlee13}. Here, ``spin'' is actually pseudospin describing one of the two atomic hyperfine states contributing to the pair formation. Using this thermodynamic quantity, we can examine to what extent the spin degrees of freedom are active in the BCS-BEC crossover region. Theoretically, the possibility of the so-called spin-gap phenomenon has been discussed in the crossover region near the superfluid phase transition temperature $T_{\rm c}$\cite{Palestini,tkashimura12,Enss,Wlazlowski,Tajima}, where the spin susceptibility is anomalously suppressed by preformed spin-singlet Cooper pairs. Since preformed Cooper pairs also cause the pseudogap phenomenon\cite{stsuchiya09,stsuchiya09b,Chen2,rwatanabe10,rwatanabe10b} (where the single-particle density of states exhibits a gap-like structure even in the normal state), the spin-gap phenomenon and pseudogap phenomenon are deeply related to each other in the cold Fermi gas system.
\par
So far, the spin susceptibility has theoretically been discussed in a uniform Fermi gas\cite{tkashimura12,Enss,Wlazlowski,Tajima}, although a real ultracold Fermi gas is always prepared in a trap potential. In this paper, thus, taking this realistic situation into account, we study how spatially inhomogeneous pairing fluctuations affect the spin-gap phenomenon in a trapped unitary Fermi gas. For this purpose, we employ the extended $T$-matrix approximation (ETMA) developed in the uniform system\cite{tkashimura12,Tajima}, to include effects of a harmonic trap within the local density approximation (LDA)\cite{stsuchiya09b,rwatanabe10b}. In a uniform Fermi gas, it has been shown that ETMA  correctly describes the BCS-BEC crossover behavior of the spin susceptibility\cite{tkashimura12,Tajima}, which makes us expect that this strong-coupling theory is also valid for the trapped case. We briefly note that the ordinary $T$-matrix approximation\cite{stsuchiya09,rwatanabe10,Strinati}, as well as the strong-coupling theory developed by Nozi\`eres and Schmitt-Rink\cite{NSR,SadeMelo}, are known to unphysically give negative spin susceptibility in the BCS-BEC crossover region, although these theories have successfully explained various many-body phenomena in the BCS-BEC crossover region. Using the combined ETMA with LDA, we calculate the local spin susceptibility $\chi(T,r)$ in the normal state near $T_{\rm c}$. Throughout this paper, we take $\hbar=k_{\rm B}=1$, for simplicity.
\par
\section{Formulation}
\par
We consider a two-component Fermi gas, described by the BCS Hamiltonian,
\begin{equation}
\label{eq2-1}
H=\sum_{\mbf{p},\s}\xi_{\mbf{p},\s}c_{\mbf{p},\s}^{\dag}c_{\mbf{p},\s}
-U\sum_{\mbf{p},\mbf{p'},\mbf{q}}c_{\mbf{p}+\frac{\mbf{q}}{2},\up}^{\dag}c_{-\mbf{p}+\frac{\mbf{q}}{2},\down}^{\dag}c_{-\mbf{p'}+\frac{\mbf{q}}{2},\down}c_{\mbf{p'}+\frac{\mbf{q}}{2},\up},
\end{equation} 
where $c^{\dag}_{\mbf{p},\s}$ is a creation operator of a Fermi atom with pseudospin $\s=\up,\down$. $\xi_{{\mbf p},\sigma}=p^2/(2m)-\mu-\sigma h$ is the kinetic energy in the $\sigma$-spin component, which is measured from the Fermi chemical potential $\mu$, where $m$ is an atomic mass, and $h$ is an infinitesimally small effective magnetic field to calculate the spin susceptibility. The pairing interaction $-U$ is assumed to be tunable. The unitarity limit (which we are dealing with in this paper) is characterized by the vanishing inverse $s$-wave scattering length ($a_s^{-1}=0$), which is related to the interaction strength $-U$ as
\begin{equation}
\label{eq2-2}
\frac{4\pi a_{s}}{m}=-\frac{U}{1-U\sum_{\mbf{p}}^{p_{\rm c}}\frac{m}{p^{2}}},
\end{equation}  
where $p_{\rm c}$ is a cut-off momentum.
\par
\begin{figure}[t]
\begin{center}
\includegraphics[width=6cm]{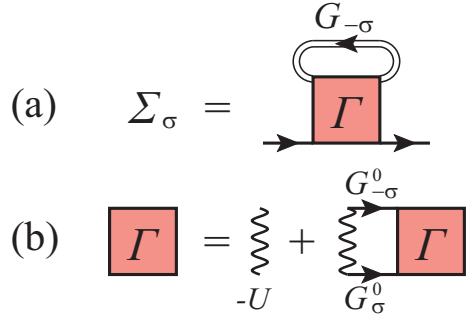}
\end{center}
\caption{(Color online) (a) Self-energy correction $\Sigma_{\s}(\mbf{p},i\omega_{n},r)$ in the extended $T$-matrix approximation (ETMA). (b) Particle-particle scattering matrix $\Gamma(\mbf{q},i\nu_n,r)$. The solid line and double-solid line represent the bare Green's function $G_\sigma^0$ and the ETMA Green's function $G_\sigma$, respectively. The wavy line describes the pairing interaction $-U$.}
\label{fig1}
\end{figure}

In LDA, effects of a harmonic trap potential $V(r)=m\Omega^2r^2/2$ can be conveniently incorporated into the theory by simply replacing the chemical potential $\mu_\sigma=\mu +\sigma h$ with the position-dependent one $\mu_\sigma(r)=\mu_\sigma-V(r)$\cite{stsuchiya09b,rwatanabe10b}, where $\Omega$ is a trap frequency. The LDA single-particle thermal Green's function then has the form,
\begin{equation}
\label{eq2-3}
G_{\s}(\mbf{p},i\omega_{n},r)=\frac{1}{i\omega_{n}-\xi_{\mbf{p},\s}(r)-\Sigma_{\s}(\mbf{p},i\omega_{n},r)},
\end{equation}
where $\omega_{n}$ is the fermion Matsubara frequency, and $\xi_{\mbf{p},\s}(r)=p^2/(2m)-\mu_\sigma(r)$. The LDA self-energy $\Sigma_{\s}(\mbf{p},i\omega_{n},r)$ describes fluctuation corrections to single-particle Fermi excitations. In ETMA, it is diagrammatically described as Fig. \ref{fig1}, which gives,
\begin{equation}
\label{eq2-4}
\Sigma_{\s}(\mbf{p},i\omega_{n},r)=T\sum_{\mbf{q},i\nu_{n}}\Gamma(\mbf{q},i\nu_{n},r)G_{-\s}(\mbf{q}-\mbf{p},i\nu_{n}-i\omega_{n},r).
\end{equation} 
Here, $\nu_{n}$ is the boson Matsubara frequency, and $-\sigma$ means the opposite component to $\sigma$-spin. We briefly note that the ordinary $T$-matrix approximation\cite{stsuchiya09,rwatanabe10,Strinati} is immediately reproduced by simply replacing the ETMA Green's function $G_{-\s}$ in Eq. (\ref{eq2-4}) with the bare one,
\begin{equation}
\label{eq2-3b}
G_{-\s}^0(\mbf{p},i\omega_{n},r)=\frac{1}{i\omega_{n}-\xi_{\mbf{p},-\s}(r)}.
\end{equation}
In Eq. (\ref{eq2-4}), $\Gamma(\mbf{q},i\nu_n,r)$ is the particle-particle scattering matrix, given by
\begin{equation}
\label{eq2-5}
\Gamma(\mbf{q},i\nu_{n},r)=\frac{-U}{1-U\Pi(\mbf{q},i\nu_{n},r)},
\end{equation}
where 
\begin{equation}
\label{eq2-6}
\Pi(\mbf{q},i\nu_{n},r)=T\sum_{\mbf{p},i\omega_{n}}G^{0}_{\up}(\mbf{p}+\mbf{q},i\omega_{n}+i\nu_{n},r)G^{0}_{\down}(-\mbf{p},-i\omega_{n},r)
\end{equation}
is the lowest-order pair-correlation function, describing fluctuations in the Cooper channel.
\par
\par
In the present formalism, the superfluid phase transition temperature $T_{\rm c}$ is determined from the condition that the Thouless criterion is satisfied at the trap center as $\Gamma(\mbf{q}=\mbf{0},i\nu_{n}=0,r=0)^{-1}=0$\cite{stsuchiya09b,rwatanabe10b}. As usual, we solve this equation, together with the equation for the total number $N$ of Fermi atoms, given by 
\begin{equation}
N=\sum_{\s}\int d\mbf{r}n_{\s}(r),
\label{eq2-6b}
\end{equation}
to self-consistently determine $T_{\rm c}$ and $\mu$. Here,
\begin{equation}
n_{\s}(r)=T\sum_{\mbf{p},i\omega_{n}}G_{\s}(\mbf{p},i\omega_{n},r)
\label{eq2-6c}
\end{equation}
is the local number density of Fermi atoms with $\sigma$ spin. Above $T_{\rm c}$, we only solve the LDA number equation (\ref{eq2-6b}), to determine the chemical potential $\mu$.
\par
The local spin susceptibility $\chi(T,r)$ is calculated from,
\begin{equation}
\label{eq2-7}
\chi(T,r)=\frac{\partial [n_{\up}(r)-n_{\down}(r)]}{\partial h}=\lim_{h\rightarrow 0}\frac{n_{\up}(r)-n_{\down}(r)}{h}.
\label{eq2-6d}
\end{equation}
In this paper, we numerically evaluate Eq. (\ref{eq2-6d}), by taking a small but finite value of $h$. 
\par
\begin{figure}[t]
\begin{center}
\includegraphics[width=12cm]{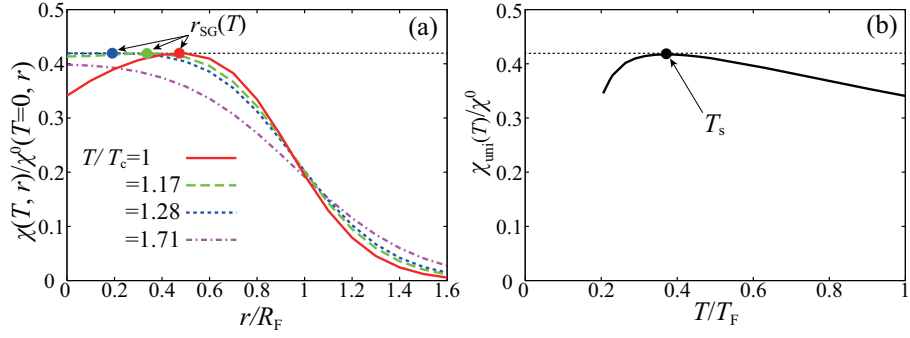}
\end{center}
\caption{(Color online) (a) Calculated local spin susceptibility $\chi(T,r)$ of a trapped unitary Fermi gas, as a function of the spatial position $r$ measured from the trap center. At each temperature, the spatial position $r_{\rm SG}(T)$ at which $\chi(T,r)$ takes a maximal value is shown as the filled circle. The horizontal dotted line shows the maximal value ($\simeq0.42$) of the scaled spin susceptibility in the case of a uniform unitary Fermi gas at $T_{\rm c}$ shown in panel (b)\cite{Tajima}. $R_{\rm F}=\sqrt{2\epsilon_{\rm F}/m}/\Omega$ is the Thomas-Fermi radius, where $\varepsilon_{\rm F}$ is the LDA Fermi energy at the trap center. $\chi^{0}(T=0,r)$ is given in Eq. (\ref{eq3-1}). (b) Spin susceptibility $\chi_{\rm uni}(T)$ in a {\it uniform} unitary Fermi gas\cite{tkashimura12,Tajima}. The filled circle shows the spin-gap temperature $T_{\rm s}\simeq0.37T_{\rm F}$ at which $\chi_{\rm uni}(T)$ takes the maximal value, $\chi_{\rm uni}(T)\simeq 0.42\chi^0$, where $\chi^{0}$ is the spin susceptibility of a free Fermi gas at $T=0$. $T_{\rm F}$ is the Fermi temperature in a uniform Fermi gas.}
\label{fig2}
\end{figure}
\par
\section{Inhomogeneous spin-gap phenomenon in a trapped unitary Fermi gas}
\par
\begin{figure}[t]
\begin{center}
\includegraphics[width=12cm]{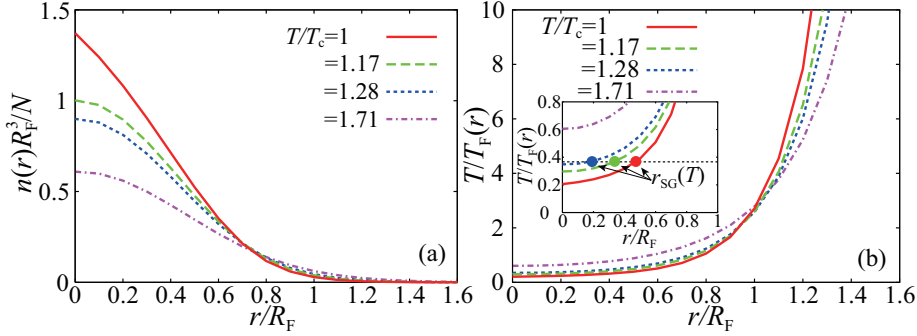}
\end{center}
\caption{(Color online) (a) Density profile $n(r)$ in a trapped ultracold Fermi gas at various temperatures. (b) Scaled temperature $T/T_{\rm F}(r)$, as a function of $r$. The inset shows $T/T_{\rm F}(r)$ magnified around the trap center, where $r_{\rm SG}(T)$ is the peak position of $\chi(T,r)/\chi^0(T=0,r)$ in Fig. \ref{fig2}(a). The horizontal dotted line in the inset shows the spin-gap temperature $T_{\rm s}/T_{\rm F}\simeq 0.37$ in a uniform Fermi gas at the unitarity.}
\label{fig3}
\end{figure}

Figure \ref{fig2}(a) shows the local spin susceptibility $\chi(T,r)$ in a trapped unitary Fermi gas above $T_{\rm c}$. Here, $\chi(T,r)$ is normalized by the zero-temperature spin susceptibility $\chi^0(T=0,r)$ in an assumed uniform free Fermi gas with the particle density being equal to the density at $r$ in the trapped case, given by
\begin{equation}
\label{eq3-1}
\chi^{0}(T=0,r)=3m(3\pi^{2})^{-\frac{2}{3}}n(r)^{\frac{1}{3}},
\end{equation}
where $n(r)=n_{\up}(r)+n_{\down}(r)$. Since the density profile monotonically decreases as one goes away from the trap center (See Fig. \ref{fig3}(a).), pairing fluctuations become weak around the edge of the gas cloud even at $T_{\rm c}$. On the other hand, atoms feel a high scaled-temperature $T/T_{\rm F}(r)$ around the edge of the gas cloud, because the LDA local Fermi temperature,
\begin{equation}
T_{\rm F}(r)=[3\pi^2n(r)]^{2/3}/2m,
\label{eq.app}
\end{equation}
is low in the low-density region. (See Fig. \ref{fig3}(b).) As a result, the local spin susceptibility $\chi(T,r)$ is suppressed thermally around the edge of the gas cloud, as in the case of a simple free Fermi gas at high temperatures. Thus, one has $\chi(T.r)/\chi^0(T=0,r)\ll 1$ in this spatial region, as seen in Fig. \ref{fig2}(a). This ordinary thermal effect becomes weak, as one approaches the trap center, because of the decrease of the scaled temperature $T/T_{\rm F}(r)$, as shown in Fig. \ref{fig3}(b). As a result, $\chi(T,r)/\chi^0(T=0,r)$ increases, as one approaches the trap center from the outer region of the gas cloud.
\par
However, Fig. \ref{fig2}(a) shows that the scaled spin susceptibility $\chi(T=T_{\rm c},r)/\chi^0(T=0,r)$ is suppressed in the vicinity of the trap center, $r\lesssim 0.46R_{\rm F}$ (where $R_{\rm F}$ is the Thomas Fermi radius), in spite of the fact that the scaled temperature $T/T_{\rm F}(r)$ still decreases with decreasing $r$ in this spatial region (because of the monotonic spatial variation of the density profile shown in Fig. \ref{fig3}(a)). Thus, this suppression is not due to the simple thermal effect, but is considered as the spin-gap phenomenon originating from strong pairing fluctuations enhanced in the trap center near $T_{\rm c}$. Indeed, in the spatial region $r\le r_{\rm SG}(T)$, where $r_{\rm SG}(T)$ is the position at which $\chi(T,r)/\chi^0(T=0,r)$ takes a maximal value, $\chi(T,r)/\chi^0(T=0,r)$ is found to {\it increase} with increasing the temperature. While this temperature dependence is opposite to the case of a uniform free Fermi gas (where the spin susceptibility monotonically {\it decreases} with an increase of the temperature), it is consistent with the temperature dependence of the spin susceptibility in the spin-gap regime ($T\le T_{\rm s}$) of a uniform Fermi gas\cite{tkashimura12,Tajima}. (See Fig. \ref{fig2}(b).) As shown in Fig. \ref{fig2}(a), the spatial region, $r\le r_{\rm SG}(T)$, becomes narrower at higher temperatures, to eventually vanish at $T\simeq 1.33T_{\rm c}$, reflecting the weakening of pairing fluctuations.
\par
\par
A uniform Fermi gas at the unitarity is known to exhibit the so-called universal thermodynamics\cite{Hu,Horikoshi,Ku}, where the Fermi energy $\varepsilon_{\rm F}$ (or equivalently the Fermi temperature $T_{\rm F}$) is the unique energy scale, because of the vanishing inverse scattering length $a_s^{-1}=0$. In the present trapped case, the scaled local spin susceptibility in LDA is expected to behave as,
\begin{equation}
{\chi(T,r) \over \chi^0(T=0,r)}=X(T/T_{\rm F}(r)).
\label{eq3-1b}
\end{equation}
The same universal function $X(x)$ in Eq. (\ref{eq3-1b}) is also expected in the uniform case as
\begin{equation}
{\chi_{\rm uni}(T) \over \chi^0}=X(T/T_{\rm F}),
\label{eq3-1c}
\end{equation}
where $\chi_{\rm uni}$ is the spin susceptibility in a uniform unitary Fermi gas, and $\chi^0$ is the zero-temperature susceptibility in a uniform free Fermi gas. $T_{\rm F}$ is the Fermi temperature in a uniform free Fermi gas. Using the relation between Eqs. (\ref{eq3-1b}) and (\ref{eq3-1c}), together with the fact that the scaled temperature $T/T_{\rm F}(r)$ is related to the spatial position through Eq. (\ref{eq.app}), we can relate the spatial variation of $\chi(T,r)/\chi^0(T=0,r)$ in Fig. \ref{fig2}(a) to the temperature dependence of $\chi_{\rm uni}(T)/\chi^0$ in Fig. \ref{fig2}(b). Indeed, the maximal value $\chi_{\rm uni}/\chi^0\simeq 0.42$ at the spin gap temperature $T_{\rm s}/T_{\rm F}\simeq 0.37$ in a uniform unitary Fermi gas (Fig. \ref{fig2}(b)) just equals the peak value of $\chi(T,r)/\chi^0(r,T=0)$ at $r=r_{\rm SG}(T)$ in the trapped case (Fig. \ref{fig2}(b)), and the latter result is independent of the value of $T$. In addition, the inset in Fig. \ref{fig3}(b) shows that the local scaled temperature $T/T_{\rm F}(r=r_{\rm SG}(T))$ in the trapped case always equals the spin gap temperature $T_{\rm s}/T_{\rm F}\simeq0.37$ obtained in the uniform case. These universal results indicate that the observations of the spatial variation of the spin susceptibility $\chi(T,r)$, as well as the density profile $n(r)$, in a trapped Fermi gas at the unitarity enable us to evaluate the spin-gap temperature $T_{\rm s}$ in a {\it uniform} unitary Fermi gas.
\par
In this regard, we briefly note that the relation between a uniform Fermi gas and a trapped one become complicated when $a_s^{-1}\ne 0$. In this case, the LDA spin susceptibility $\chi(T,r)$ in a trap also depends on $(p_{\rm F}(r)a_s)^{-1}$ in addition to $T/T_{\rm F}(r)$, where $p_{\rm F}(r)=[3\pi^{2}n(r)]^{1/3}$ is the LDA local Fermi momentum. As a result, $\chi(T,r)$ is related to the spin susceptibility in a uniform Fermi gas, not only at various scaled temperatures $T/T_{\rm F}$, but also at various interaction strengths $(p_{\rm F}a_s)^{-1}$, where $p_{\rm F}$ is the Fermi momentum in a uniform Fermi gas. 
\par
\section{Summary}
\par
To summarize, we have discussed magnetic properties of a unitary Fermi gas in a harmonic potential above $T_{\rm c}$. Including strong pairing fluctuations within the framework of the extended $T$-matrix approximation (ETMA), as well as effects of a harmonic trap within the local density approximation (LDA), we showed that, near $T_{\rm c}$, the local spin susceptibility is anomalously suppressed in the trap center due to the formation of preformed singlet Cooper pairs. The spatial region where this spin-gap phenomenon occurs becomes wide with decreasing the temperature. We also confirmed that the so-called universal thermodynamics hold for the spin susceptibility. We pointed out that, using this, we can determine the spin-gap temperature $T_{\rm s}$ in a uniform unitary Fermi from the observation of the spatial variation of the local spin susceptibility in the trapped case.
\par
In this paper, we have treated effects of a harmonic trap within LDA, where spatial correlations are completely ignored. In addition, the present analyses is restricted to the unitarity limit. Improving these issues remains as our future problems. Since a real ultracold Fermi gas is always trapped in a harmonic potential, our results would be useful for the study of how the spatial inhomogeneity affects thermodynamic properties of this system in the BCS-BEC crossover region, as well as how to observe the spin gap temperature $T_{\rm s}$ in a unitary Fermi gas. 
\par
\par
\begin{acknowledgements}
We would like to thank T. Kashimura, R. Watanabe, D. Inotani and P. van Wyk for useful discussions. This work was supported by the KiPAS project in Keio university. H.T. and R.H. were supported by the Japan Society for the Promotion of Science. Y.O was supported by Grant-in-Aid for Scientific Research from MEXT and JSPS in Japan (No.25400418, No.15H00840).
\end{acknowledgements}


\end{document}